\pdfoutput=1

\documentclass[11pt]{article}

\usepackage[final]{acl}

\usepackage{times}
\usepackage{latexsym}
\usepackage{amsmath}
\usepackage{amssymb}

\usepackage{tabularx}
\usepackage{hyperref}
\usepackage{array}

\usepackage{xcolor}

\usepackage{enumitem}

\usepackage[T1]{fontenc}

\usepackage[utf8]{inputenc}

\usepackage{microtype}

\usepackage{inconsolata}

\usepackage{graphicx}
\usepackage{booktabs}
\usepackage{multirow}
\usepackage{makecell}

\newcolumntype{C}[1]{>{\centering\arraybackslash}m{#1}}
\newcolumntype{P}[1]{>{\raggedright\arraybackslash}m{#1}}
\newcolumntype{L}{>{\raggedright\arraybackslash}X}

\title{Do Retrieval Ranking Models Capture Event Order? \\ A Study of Timeline-Constrained Retrieval}

\author{
    Alon Eirew$^{1}$ \quad Lidan Avisar$^{2}$ \quad Kfir Bar$^{2}$ \quad Ido Dagan$^{1}$ \\
    $^{1}$Computer Science Department, Bar-Ilan University \\
    $^{2}$Efi Arazi School of Computer Science, Reichman University \\
    \texttt{alon.eirew@gmail.com}, \quad \texttt{kfir.bar@runi.ac.il}
}

\begin{document}
\maketitle
\begin{abstract}

Semantic retrieval models aim to match documents to a query's ``meaning'', yet it remains unclear whether they successfully encode temporal relations between events in their ranking mechanisms. We investigate this question by studying timeline-constrained document retrieval, where a query is an ordered sequence of natural-language event statements while a relevant document should match those events in a compatible temporal progression. We then test whether widely used dense retrievers and neural rerankers capture such constraints, and whether explicitly preserving temporal order improves ranking, using a controlled setting over a novel focused evaluation collection. Experiments show that state-of-the-art semantic retrievers and rerankers provide only limited improvements over lexical retrieval, suggesting that their semantic matching capabilities do not reliably capture temporal constraints. We then compare a timeline-aware method with both semantic matching and standard rerankers and find that it yields substantial improvements. This demonstrates that considering temporal order can provide a strong ranking signal in this setting, suggesting future research on improving temporal modeling in semantic retrievers.
\end{abstract}

\section{Introduction}
\label{sec:intro}

\begin{figure}[!t]
\centering
\includegraphics[width=.38\textwidth]{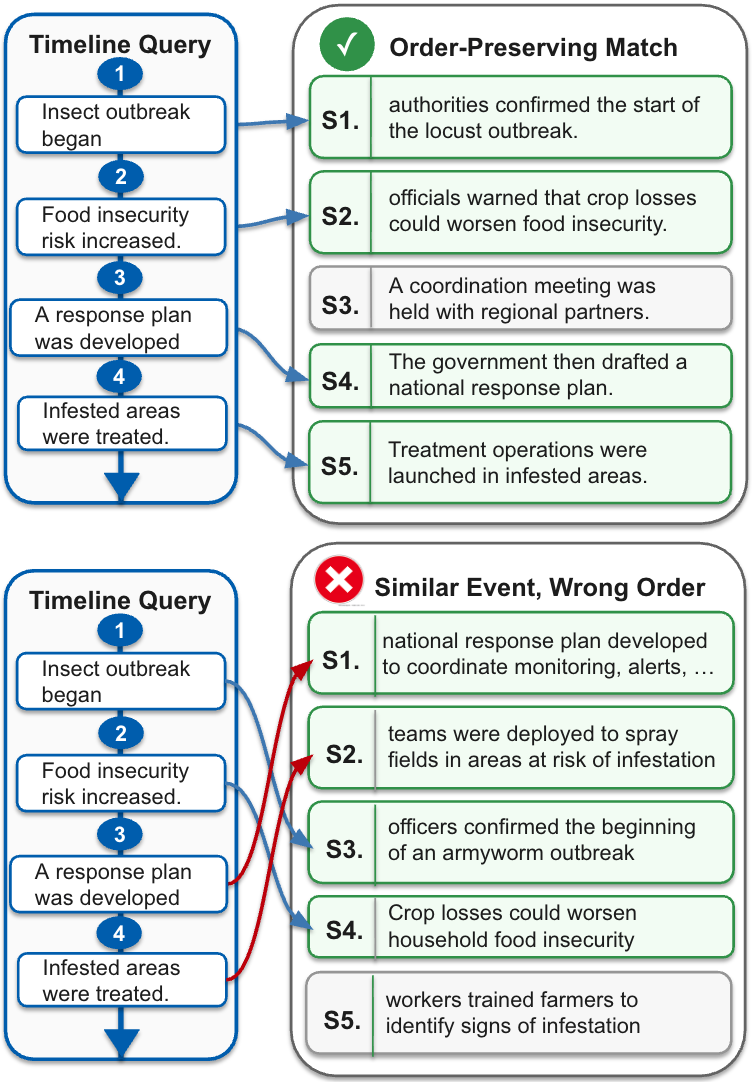}
\caption{Given a timeline query and a chronologically ordered timeline of events extracted from a candidate disaster report, relevance requires both semantic support for all query statements and consistency with their temporal order. The top candidate is relevant because its events progression aligns with all query statements in an order-preserving manner, while unrelated sentences are skipped. The bottom candidate, although supports the same events, is inconsistent with the progression specified by the query timeline.}
\label{fig:pairwise}
\end{figure}

Semantic retrieval models \citep{manning2008ir,robertson-2009-prob} aim to move beyond lexical overlap by assigning high relevance to documents that match the query with respect to its abstract concepts, and the relationships between them. In this paper, we focus on information needs that specify temporal relations between events. For example, a user may seek reports describing a natural disaster followed by infrastructure damage, population displacement, and the subsequent delivery of aid. A document that discusses all these events may be topically and semantically relevant, yet fail to satisfy the information need if its event timeline contradicts the required progression. This distinction is difficult for standard retrieval and reranking models, which primarily capture topical relevance or local semantic compatibility \cite{ye-etal-2025-sequence, han-etal-2025-temporal}.

Such timeline-constrained information needs arise across several domains. In healthcare, patient records describe trajectories of diagnoses, treatments, medications, and outcomes. Retrieving prior cases with comparable clinical courses aids prognosis and treatment planning \citep{wang-etal-2024-wonder}. In law-enforcement and investigative settings, reports of incidents such as kidnappings, assaults, or murders may be compared to identify related cases and recurring temporal patterns indicative of a common modus operandi.


In this work, we investigate whether retrieval and reranker models can capture temporal relations expressed in queries and whether explicitly modeling these relations improves retrieval performance. Our setting consists of a collection of full documents, and timeline queries expressed as ordered sequences of natural-language statements, each describing an event within a broader progression. Given such a query, the task is to rank documents according to both whether they support the specified events and whether those events unfold in a temporal progression compatible with the order expressed in the query (illustrated in Figure~\ref{fig:pairwise}). In this context, we address two research questions:

\textbf{RQ1:} How effectively do semantic retrievers and rerankers capture temporal relations between query events? We compare dense retrieval and neural reranking models with lexical retrieval. The results show that state-of-the-art retrieval (and reranking) models provide only limited improvements over lexical retrieval, indicating that their semantic matching capabilities do not effectively capture the temporal constraints that determine relevance.


\textbf{RQ2:} Does preserving temporal order improve reranking performance? We introduce an order-preserving alignment method and show that enforcing temporal order substantially outperforms standard reranking models.

To support this investigation, we make the following contributions:\footnote{All experimental code, LLM prompts, and data will be made publicly available.}

\begin{itemize}[leftmargin=*,itemsep=0pt,topsep=0pt]
    \item We formulate a controlled setting for timeline-aware text retrieval and construct a focused evaluation collection using a decomposed protocol that separately assesses statement-level semantic support and temporal consistency (§\ref{sec:task_benchmark}).
    \item We systematically evaluate and compare 
    lexical and dense retrieval, as well as neural reranking, under timeline-aware conditions, thereby isolating the contribution of explicitly preserving temporal order (\S\ref{sec:method})
    \item We show that order-preserving alignment significantly outperforms both standard  retrievers and rerankers, demonstrating the specific contribution of explicit temporal-order preservation (\S\ref{sec:experiments}).
\end{itemize}

\section{Related Work}
\label{sec:related_work}

Neural retrieval and reranking models aim to capture semantic compatibility beyond lexical overlap. However, strong performance on conventional retrieval benchmarks does not necessarily imply sensitivity to the specific semantic relations that determine relevance. Controlled evaluations have shown that neural retrieval models can struggle with constraints expressed through negation or natural-language instructions \citep{weller-etal-2024-nevir,weller-etal-2025-followir}. More closely related to our setting, \citet{ye-etal-2025-sequence} find that general retrieval models often neglect the sequence structure of procedural documents. These findings motivate our investigation of temporal relations as a focused test of whether semantic retrieval models capture relational constraints expressed in queries.

Temporal information has long been studied in information retrieval. Prior temporal-IR approaches incorporate signals such as publication time, recency, temporal expressions, and a query's temporal intent \citep{anand2012temporal, campos2014survey,han-etal-2025-temporal}. These approaches primarily model \emph{when} information or documents are relevant. In contrast, we examine whether the internal progression of events described by a document is compatible with the temporal relations expressed in a natural-language query.

Event timelines have also been studied in timeline extraction and temporal-relation modeling \citep{rogers-etal-2024-narrativetime, wang-etal-2024-wonder, eirew-etal-2025-beyond, homburger-bar-2025-large}, timeline summarization \citep{pratapa-etal-2023-background,hu-etal-2024-moments,zhang-etal-2025-dtels}, and temporal reasoning and question answering \citep{tan-etal-2023-timelineqa,luo-etal-2025-etrqa}. These tasks generally assume that the relevant source documents have already been identified and seek to extract, organize, summarize, or reason over their temporal information, rather than use event order to rank documents.

The closest line of research combines retrieval with timeline or event-sequence modeling. \citet{chieu-2004-timeline} retrieve query-relevant event sentences and organize them into a timeline. \citet{gupta-etal-2022-learning,gupta-etal-2025-retrieving} retrieve structured continuous-time sequences of timestamped events, while \citet{liu-quan-2025-retrieval} retrieve such sequences from textual descriptions. These settings retrieve individual event sentences or structured timestamped sequences. We instead rank full, unstructured documents according to both semantic support for the query events and consistency with their specified temporal progression.

\section{Study Setting and Evaluation Dataset}
\label{sec:task_benchmark}

We introduce a controlled evaluation framework that separates semantic support for individual query events from compatibility with their specified temporal progression. We formalize timeline-constrained relevance (\S\ref{sec:task_definition}), describe the relevance assessment protocol used to construct and validate the collection (\S\ref{sec:decomposed_assessment}), and present the collection and its validation (\S\ref{sec:benchmark}--\ref{sec:benchmark_stats}).

\subsection{Timeline-Constrained Retrieval Setting}
\label{sec:task_definition}

To study the contribution of temporal order to document ranking, we represent a timeline query as an ordered sequence of natural-language statements, $q=(q_1,\ldots,q_m)$, where each $q_i$ describing an event—a real-world occurrence or change of state, such as infrastructure damage, displacement, or aid delivery. The query statements are arranged according to the required chronological progression. Given a document collection $D=\{d_1,\ldots,d_n\}$, the objective is to rank the candidate documents according to how well the document internal event timelines align with the progression specified by $q$.

Importantly, temporal-order consistency is defined over the chronological relations between events, rather than the textual order in which those events are mentioned. Documents frequently describe events non-chronologically. For example, in the sentence ``The \textit{investigation} found that food \textit{poisoning} was the cause of the \textit{sickness},'' the investigation is mentioned before the poisoning and sickness, although both occurred earlier in time. Therefore, a relevant document need not provide evidence for $q_1,\ldots,q_m$ in the same textual order as the query statements. Instead, the \textit{timeline} extracted from the document---in our example, \textit{poisoning} $\rightarrow$ \textit{sickness} $\rightarrow$ \textit{investigation}---must support the query timeline (illustrated in Figure~\ref{fig:pairwise}).

Following standard information-retrieval evaluation \citep{manning2008ir}, we define binary relevance. A query--document pair is considered \textsc{Relevant} if it satisfies two conditions: 

\textbf{Semantic support.} every query statement is supported by evidence somewhere in the document under a textual entailment criterion \citep{dagan2006rte}. In our setting, evidence from the document serves as the premise, and a query statement serves as the hypothesis. The evidence \textsc{Supports} the query statement if it entails that the statement is true, and is labeled \textsc{NotSupported} if it either entails that the statement is false (i.e., contradicts), or if it does not provide sufficient information to determine whether the statement is true or false (i.e., neutral) (illustrated in Figure~\ref{fig:method} (2b.)).

\textbf{Temporal consistency.} the temporal ordering (timeline) of the supported events in the document is consistent with their ordering in the query. We adopt a broad definition of temporal consistency: for any two consecutive query statements $q_i$ and $q_{i+1}$, a document is considered temporally consistent if the corresponding evidence appears in the same temporal order, simultaneously, or when their temporal relation is ambiguous or cannot be determined from the provided context. Only documents in which the evidence clearly contradicts the query ordering, that is, $q_{i+1}$ precedes $q_i$, are considered temporally inconsistent.

This definition differ from topical retrieval: a document may mention the same types of events as the query, but still be irrelevant if the events appear in an incompatible temporal order (Figure~\ref{fig:pairwise}).

\subsection{Query--Document Relevance Assessment}
\label{sec:decomposed_assessment}

\begin{figure}[!t]
\centering
\includegraphics[width=.46\textwidth]{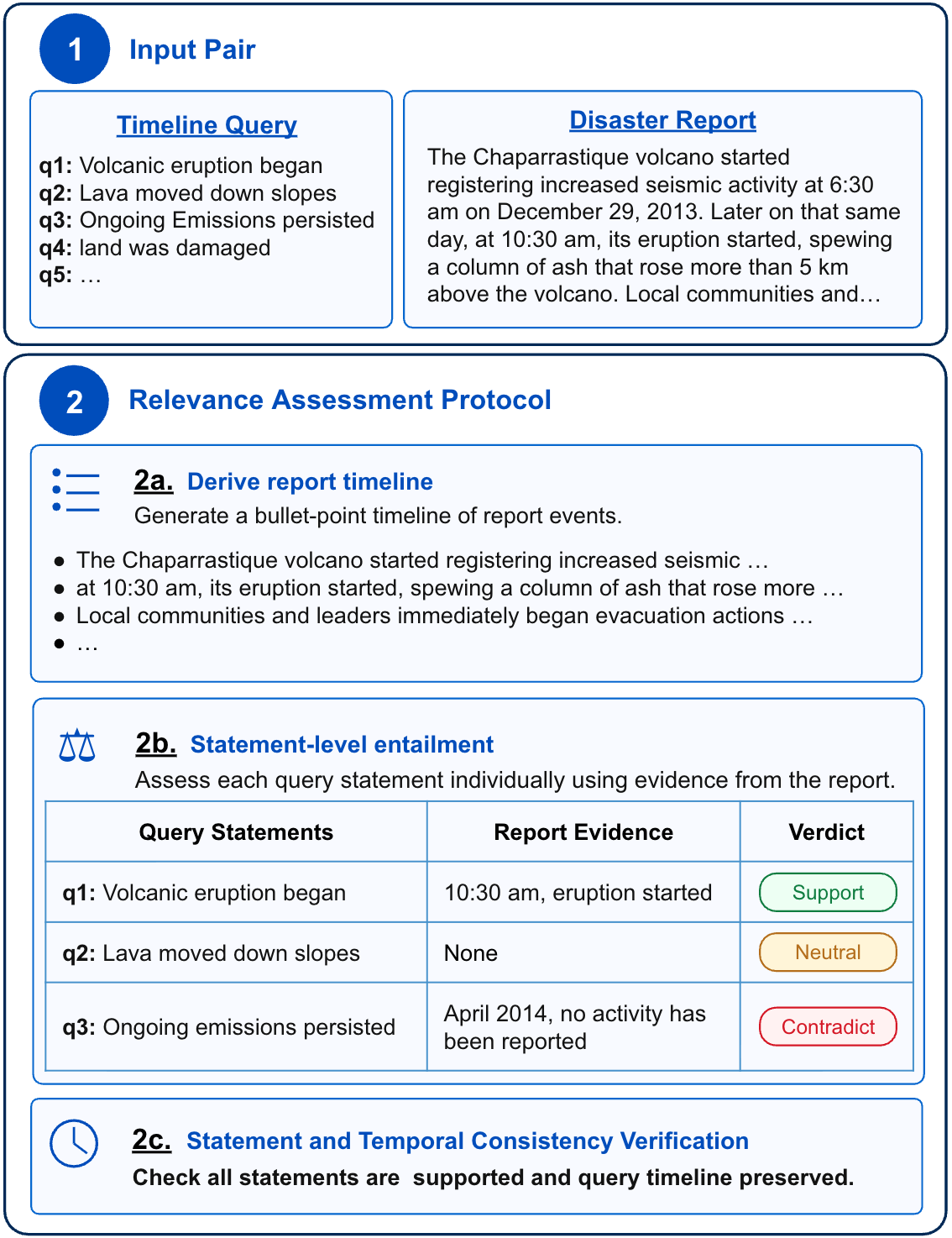}
\caption{
Query–document relevance assessment protocol. The protocol is detailed in §\ref{sec:decomposed_assessment}. In this example, we use ``...'' to indicate sentence fragments trimmed to fit within the figure.}
\label{fig:method}
\end{figure}

Assessing timeline relevance is hard because it couples two conditions: statement-level semantic support and global order consistency.
We therefore operationalize the relevance definition through a query--document assessment protocol, illustrated in Figure~\ref{fig:method}.
Given a timeline query and a candidate document, we first derive a natural-language timeline of the events described in the document. 
We then assess each query statement independently under an entailment-style framework: for each statement, we identify the relevant text segment from the derived document timeline as evidence, and assign a statement-level label indicating whether the evidence supports, is neutral toward, or contradicts the statement.
Finally, determine whether the chronological relations among the supported events are compatible with the event progression specified by the query.

We use this assessment protocol in two ways. 
We apply it with a strong LLM to label our evaluation collection  (§\ref{sec:benchmark}), and again during human validation (§\ref{sec:benchmark_stats}) to assess the quality of the generated labels.

\subsection{Evaluation Collection}
\label{sec:benchmark}

We construct an evaluation collection from disaster reports published on ReliefWeb, comprising a document corpus, a set of timeline queries, and query--document relevance judgments.\footnote{\url{https://reliefweb.int/}, a service of the United Nations Office for the Coordination of Humanitarian Affairs (OCHA) that collects and disseminates information about humanitarian crises and disasters. The dataset is licensed under CC BY 4.0.}
We select ReliefWeb because it provides a large, public collection of real-world reports describing evolving disaster events. Comparable documents from domains such as healthcare, law, and criminal investigations are difficult to obtain because of privacy, legal, and confidentiality constraints.

\paragraph{Document collection.}

Starting from 108,841 ReliefWeb documents, we use the collection metadata to identify disaster-related reports and an LLM-based filter to retain those centered on a single major disaster sequence (e.g., a storm followed by flooding, displacement, and relief efforts). We exclude regional updates and summaries of multiple unrelated events to isolate reports with a coherent event timeline and reduce ambiguity in both timeline extraction and relevance assessment (Appendix~\ref{apdx:multi_filter}). This process yields 6,145 documents suitable for timeline-aware retrieval evaluation.

\paragraph{Query generation.}

To construct the queries, we prompt GPT-5.2 \cite{openai2025gpt52} to generate bullet-point timelines from a random sample of 200 documents, 
instructing it to avoid specific identifiable entities, such as locations, dates, times, and names (the full prompt is provided in Appendix~\ref{apdx:additional}).
We manually review the generated queries and filter out those containing fewer than three timeline statements or more than ten, since in our preliminary experiments such queries tended to yield either too few or too many relevant documents. Finally, we assess the generated queries for quality and uniqueness, removing redundant and low-quality queries, including cases where event-specific information was leaked. 
The resulting collection contains 66 timeline queries.

\paragraph{Relevance judgment construction.}

To operationalize our general relevance-assessment protocol (\S\ref{sec:decomposed_assessment}; Figure~\ref{fig:method}, Part~2) for LLM-based annotation, we construct a detailed prompt for GPT-5.2 (Appendix~\ref{apdx:additional}, Figure~\ref{fig:pairs_prompt}). The prompt first instructs the model to derive an event timeline from each document using instructions adapted from \citet{eirew-etal-2025-beyond}. It then assesses each query statement $q_i$ against the events in the derived document timeline. For each $q_i$, the model produces a tuple $(\ell_i,e_i,r_i)$, where $\ell_i$ is a statement-level entailment label, $e_i$ is cited evidence from the document, and $r_i$ is an explanation for the $\ell_i$ judgment. Based on these statement-level assessments, the model assigns an overall relevance label $y(q,d)$ to the candidate query--document pair by determining whether all query statements are supported by document evidence and whether the document timeline is temporally consistent with the query timeline.

To verify that the LLM output is complete and grounded in the source document, we parse each response into a structured object and check that: (i) exactly one assessment is provided for every $q_i$; (ii) no assessment is introduced for a statement that does not appear in $q$; and (iii) every cited evidence sentence $e_i$ can be matched to the original document $d$.\footnote{Evidence sentences are verified using fuzzy matching implemented with RapidFuzz \citep{rapidfuzz}.} If any check fails, we regenerate the assessment for that query--document pair. Appendix~\ref{apdx:additional}, Figure~\ref{fig:hard_neg1}--\ref{fig:hard_neg2}, provides a qualitative hard-negative example.

\subsection{Collection Statistics and Validation}
\label{sec:benchmark_stats}

Overall, the evaluation collection comprises 66 timeline queries: 50 test queries and 16 development queries, with an average of six statements per query. After excluding the 66 source documents used to generate the queries, the document corpus contains 6,079 ReliefWeb disaster reports spanning 15 disaster categories. The gold relevance judgments comprise 234 relevant and 400,980 irrelevant query--document pairs, corresponding to an average of 3.5 relevant documents per query. The documents contain an average of 500 words.\footnote{Additional dataset statistics are provided in Appendices~\ref{apdx:disaster_categ} and~\ref{apdx:additional}.}

To assess the validity of our LLM-generated labels, we conducted the following assessments:
For statement-level consistency, we randomly sampled 70 query--document pairs comprising 426 query statements. We then regenerated the annotations using the same relevance assessment protocol (§\ref{sec:decomposed_assessment}) and evaluated model consistency. The model changed its statement-level verdict (Figure~\ref{fig:method} (2b.)) in only 3\% of the cases.
We further evaluated the temporal-order consistency of the LLM by modifying one timeline statement in 25 queries. We ensured that each modified timeline remained plausible and could represent a realistic alternative unfolding of a similar event (for example, by replacing query statement (2) with query statement (3) in Figure~\ref{fig:pairwise}). We then re-evaluated the originally relevant reports, yielding 70 re-examined pairs. The model reversed its final relevance verdict in 86\% of these cases, confirming that the order in which sentences are presented has a meaningful impact. 
We analyze the unchanged cases in Appendices~\ref{apdx:temp_order_consist}, and~\ref{apdx:additional}; most involve ambiguous temporal relations, which are treated as consistent under our definition (§\ref{sec:task_definition}).

Finally, two authors independently annotated 50 randomly selected query--document pairs for validation, using the relevance assessment protocol (§\ref{sec:decomposed_assessment}), including 25 positive examples and 25 hard negatives, where all query statements were supported but the model judged the pair as \textsc{Irrelevant} due to a temporal-order mismatch with the query. The authors showed high inter-annotator agreement, with Cohen's $\kappa=0.83$. Agreement with the LLM-generated labels was also substantial, with Cohen's $\kappa=0.8$ for each annotator compared against the LLM labels.

\section{Temporal Order in Document Ranking}
\label{sec:method}

\begin{figure}[!t]
\centering
\includegraphics[width=.35\textwidth]{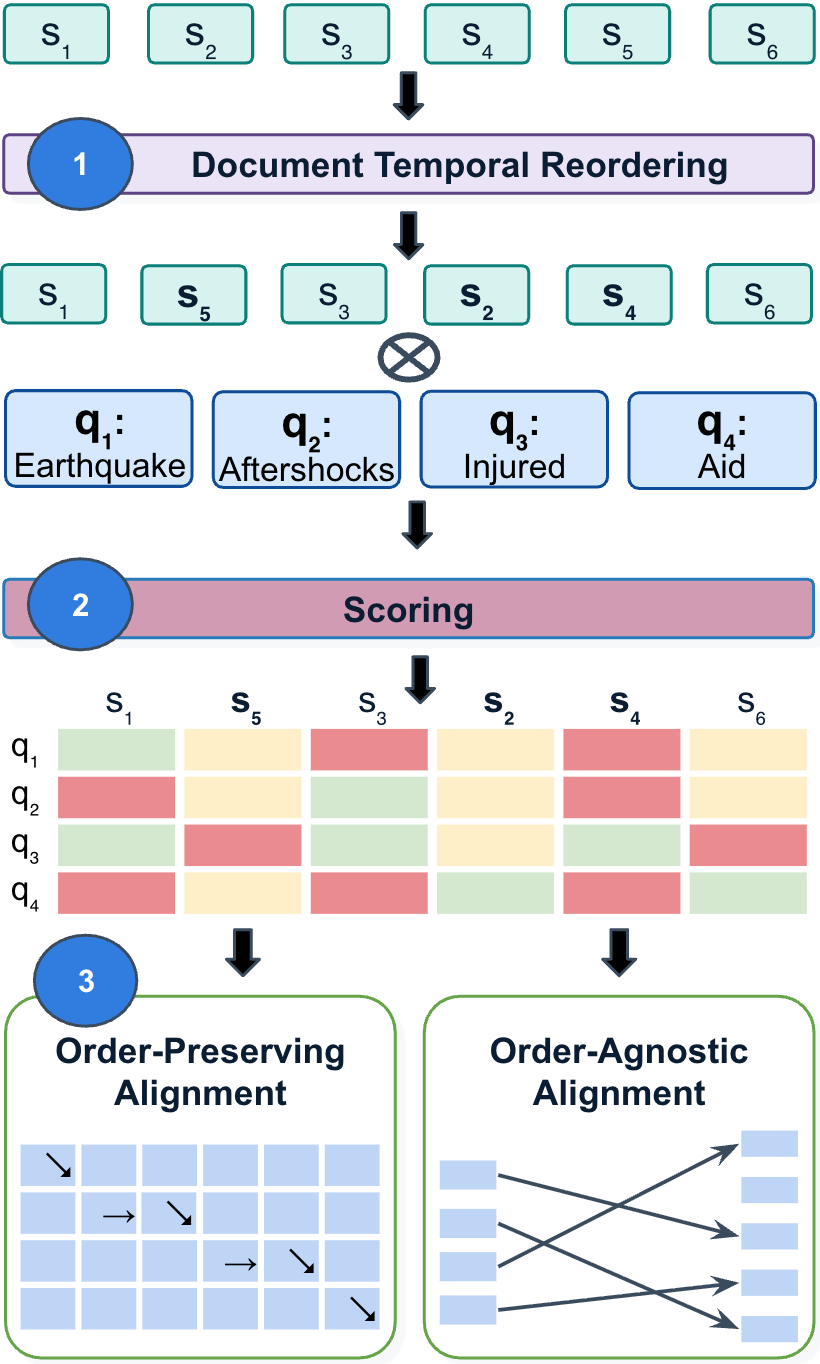}
\caption{The order-preserving and order-agnostic statement--sentence alignment framework for timeline-aware reranking (§\ref{sec:monotonic_alignment}).}
\label{fig:model}
\end{figure}

Our framework compares two alignment conditions that share the same statement--sentence compatibility scores and differ only in whether matches must preserve the query progression. Because textual document narratives
may not reflect event chronology, we first temporally order each document (\S\ref{sec:document_ordering}), compute pairwise compatibility scores (\S\ref{sec:local_matching}), and apply either order-preserving (\S\ref{sec:monotonic_alignment}) or order-agnostic (\S\ref{sec:non_monotonic}) alignment (Figure~\ref{fig:model}).

\subsection{Chronological Document Representation}
\label{sec:document_ordering}

In an offline preprocessing step, we split each document into sentences and prompt an LLM to order them chronologically by the events they describe (Step 1 in Figure~\ref{fig:model}).\footnote{The total cost of this step using GPT-5.2 was under \$50.} Only the ordering changes; the sentence text is untouched. The resulting sequence is stored and reused at ranking time, and since the step is query-independent, it runs once per document.\footnote{Further detailed in Appendix~\ref{apdx:doc_chrono_order}.}

\subsection{Statement--Sentence Compatibility}
\label{sec:local_matching}

Given a timeline query $q=(q_1,\dots,q_m)$ and a candidate document now represented as a temporally ordered sentence sequence $d=(s_1,\dots,s_n)$, we first compute a sentence-level similarity matrix between the query and the candidate document. For each query statement $q_i$ and document sentence $s_j$, we use a pretrained sentence-pair scoring algorithm \citep{wang2020minilm, chen-etal-2024-m3} to assign a semantic compatibility score 
$S_{ij}=\phi(q_i,s_j),$
where $\phi$ is the scoring function (a cross-encoder in our experiments) and $S \in \mathbb{R}^{m \times n}$ (Step 2 in Figure~\ref{fig:model}). Intuitively, $S_{ij}$ estimates how well document sentence $s_j$ is semantically similar to query statement $q_i$.

We next describe our order-preserving and order-agnostic alignment methods, which isolate the contribution of temporal constraints. Both methods operate over the same score matrix $S$ and differ only in whether the selected statement--sentence matches must preserve the temporal progression specified by the query.

\subsection{Order-Preserving Alignment}
\label{sec:monotonic_alignment}

Given the score matrix $S$, we search for the highest-scoring monotonic alignment between the query statements and the temporally ordered document sentences. Let $F(i,j)$ denote the best alignment score between the query prefix $(q_1,\dots,q_i)$ and the document prefix $(s_1,\dots,s_j)$. We use a dynamic-programming procedure inspired by classical sequence-alignment algorithms \citep{needleman1970general, smith1981}, adapted to a semi-global (ends-free) setting \citep{huang2006align,sosic2017edlib}. This formulation fits our retrieval task because a relevant document may contain additional material before, after, or between the sentences that support the query timeline.

The recurrence considers three transitions:
\setlength{\belowdisplayskip}{3pt}
\[
F(i,j) =
\max
\begin{cases}
F(i-1,j-1) + S_{ij}, \\
F(i-1,j) + g_q, \\
F(i,j-1) + g_s .
\end{cases}
\]

The first transition matches query statement $q_i$ with document sentence $s_j$. The second leaves a query statement unmatched, while the third skips a document sentence. The gap penalties $g_q$ and $g_s$ control the cost of leaving query statements or document sentences unmatched, respectively. Since skipping a query statement indicates incomplete support, it is penalized. Document sentences, however, may be skipped without penalty, as documents often contain background or intervening content. We therefore set $g_s=0$ and tune $g_q$ on the development set (see §\ref{sec:experiments}). The final alignment score, $F(m,n)$, is computed after processing the full document.

To further encourage full query coverage, we add an explicit coverage term. After computing $F(m,n)$, we recover the corresponding alignment path $P$ by backtracking from $(m,n)$. The path $P$ consists of match and gap transitions, and we let $|P_q|$ denote the number of query statements that are matched to document sentences along this path. Coverage is computed as follows:
\setlength{\belowdisplayskip}{3pt}
\[
\mathrm{cov}(q,d) = \frac{|P_q|}{m}.
\]

The final ranking score is:
\[
f(q,d) = \frac{F(m,n)}{m} + \alpha \cdot \mathrm{cov}(q,d),
\]
where $\alpha$ controls the contribution of query coverage. This score favors documents that both align well semantically and support a larger fraction of the query timeline in the correct order.\footnote{We tune $\alpha$ on the development set; Appendix~\ref{apdx:additional}, Figures~\ref{fig:gapq_trand} and~\ref{fig:alpha_trand}, illustrates the effect of different $g_q$ and $\alpha$ values on ranking performance.} Examples of the resulting alignment paths, including cases with complete and partial query coverage, are provided in Appendix~\ref{apdx:additional} Table~\ref{tab:matched_query_document_examples}).

After constructing the compatibility matrix, the order-preserving alignment runs in $O(mn)$ time; further complexity details are provided in Appendix~\ref{apdx:alignment_complexity}.

\subsection{Order-Agnostic Alignment}
\label{sec:non_monotonic}

The order-agnostic (OA) condition computes a maximum-weight
partial one-to-one bipartite matching between query statements
and document sentences using the Hungarian algorithm
\citep{Kuhn1955TheHM}. This condition captures local semantic
compatibility without constraining the order of the matched
evidence.

To permit partial matching, we augment the document side of the
bipartite graph with \(m\) dummy nodes. Matching a query
statement to a real document sentence contributes \(S_{ij}\),
whereas matching it to a dummy node contributes the query-gap
score \(g_q\). We tune \(g_q\) on the development set and use
the selected value for both OA and OP.

The two conditions use the same compatibility matrix,
partial one-to-one matching objective, query-gap score,
coverage reward, and final scoring function. They differ only
in their feasible alignments: OA permits matches in any
document order, whereas OP requires the matched document
sentences to preserve the order of the query statements.

\section{Experiments}
\label{sec:experiments}

We evaluate using the widely adopted two-stage retrieval setting \cite{khattab2020colbert}. First-stage retrievers rank the full document collection for each timeline query, after which a second-stage reranker rescores the retrieved query–document pairs. We describe the experimental setup in §\ref{sec:expr_setup} and present the results in §\ref{sec:results}.


\subsection{Experimental Setup}
\label{sec:expr_setup}

\paragraph{Query formulation.}
We evaluate several query formulations to test whether state-of-the-art semantic retrievers and rerankers can capture temporal constraints when these constraints are made more explicit. For models that support instructions, we evaluate both standard retrieval and reranking instruction configurations and temporally informed instructions that ask the model to account for the specified progression of events (prompts provided in Appendix~\ref{apdx:instruct}). Results are reported for our best effort instructions. For models without an instruction interface, we concatenate the query statements either directly (using a space delimiter) or with a temporal connective (e.g., ``then'') inserted between consecutive statements.

\paragraph{Retrievers.}
For first-stage retrieval, we evaluate several recent strong dense retrieval models, including E5-Large and E5-Large-Instruct \citep{wang2024multilingual}, BGE-M3 \citep{chen-etal-2024-m3}, and Qwen3-Embedding-0.6B \citep{zhang2025qwen3embeddingadvancingtext}. We also evaluate BM25 \citep{robertson-2009-prob},\footnote{\url{https://github.com/deepset-ai/haystack}} as a lexical candidate which achieves the highest candidate recall among the evaluated retrievers and is therefore used as the first-stage retriever for all reranking experiments detailed next. For models whose context window cannot accommodate the full document, we use the segmentation and pooling strategy described in Appendix~\ref{apdx:ret_impl_details}.

\paragraph{Neural rerankers.}
We evaluate several strong neural rerankers by rescoring the top 250 documents retrieved by BM25.\footnote{We use this cutoff because BM25 achieves 100\% Recall@$250$.} candidates returned by BM25. These include BGE-Reranker-v2-m3 \citep{chen-etal-2024-m3}, a multilingual cross-encoder; MiniLM-L-12-v2 \citep{wang2020minilm}, a lightweight cross-encoders trained on MS MARCO \citep{Campos2016MSMA}; and Qwen3-0.6B-Rerank \citep{zhang2025qwen3embeddingadvancingtext}. Each query--document pair is scored directly as a single input pair. For models whose context window cannot accommodate the full document, we use the segmentation and pooling strategy described in Appendix~\ref{apdx:rerank_impl_details}.

\paragraph{Evaluation.}
We evaluate model performance using standard information retrieval metrics \cite{manning2008ir}: MRR@$10$, Recall@$k$, and mAP@$k$. 
For end-to-end evaluation, we report Recall@$10$, Recall@$50$, mAP@$10$, and mAP@$50$. We use mAP rather than nDCG because our evaluation uses binary rather than graded relevance labels.


\begin{table}[t]
\centering
\resizebox{0.98\columnwidth}{!}{
\begin{tabular}{lccccccc}
    \toprule
    \textbf{Model} &
    \textbf{Ord.} &
    \textbf{Conn.} &
    \textbf{MRR@10} &
    \textbf{mAP@10} &
    \textbf{mAP@50} &
    \textbf{R@10} &
    \textbf{R@50} \\
    \midrule

    \multicolumn{8}{c}{\textit{Retrievers}} \\
    \midrule
    BGE-M3
        & -- & $\times$ & 0.28 & 0.16 & 0.17 & 0.30 & 0.54 \\
    BGE-M3
        & -- & $\checkmark$ & 0.27 & 0.15 & 0.17 & 0.27 & 0.52 \\
    E5-Large
        & -- & $\times$ & \textbf{0.48} & 0.25 & 0.27 & 0.31 & 0.53 \\
    E5-Large
        & -- & $\checkmark$ & 0.36 & 0.20 & 0.22 & 0.32 & 0.49 \\
    E5-Large-Inst
        & $\times$ & -- & 0.33 & 0.17 & 0.18 & 0.26 & 0.40 \\
    E5-Large-Inst
        & $\checkmark$ & -- & 0.27 & 0.15 & 0.29 & 0.28 & 0.37 \\
    Qwen3-Embed
        & $\times$ & -- & \textbf{0.48} & 0.29 & 0.31 & 0.41 & 0.65 \\
    Qwen3-Embed
        & $\checkmark$ & -- & 0.46 & 0.29 & 0.30 & 0.43 & 0.60 \\
    BM25
        & -- & -- & \textbf{0.48} & \textbf{0.35} & \textbf{0.38} & \textbf{0.49} & \textbf{0.77} \\
    \midrule
    \multicolumn{8}{c}{\textit{Rerankers}} \\
    \midrule
    Qwen3-Rerank
        & $\times$ & -- & 0.69 & 0.45 & 0.47 & 0.59 & 0.82 \\
    Qwen3-Rerank
        & $\checkmark$ & -- &
        0.65 & 0.44 & 0.46 & 0.59 & 0.79 \\
    \midrule
    BGE-Reranker
        & -- & $\times$ & 0.49 & 0.28 & 0.30 & 0.43 & 0.69 \\
    BGE-Reranker
        & -- & $\checkmark$ & 0.53 & 0.33 & 0.35 & 0.47 & 0.70 \\
    BGE-Reranker (OA)
        & -- & -- & 0.46 & 0.23 & 0.25 & 0.41 & 0.62 \\
    BGE-Reranker (OP)
        & -- & -- & \textbf{0.65} & \textbf{0.45} & \textbf{0.47} & \textbf{0.60} & \textbf{0.80} \\
    \midrule
    MiniLM-L12
        & -- & $\times$ & 0.57 & 0.32 & 0.35 & 0.48 & 0.72 \\
    MiniLM-L12
        & -- & $\checkmark$ & 0.53 & 0.30 & 0.32 & 0.45 & 0.68 \\
    MiniLM-L12 (OA)
        & -- & -- & 0.55 & 0.36 & 0.38 & 0.50 & 0.77 \\
    MiniLM-L12 (OP)
        & -- & -- &
        \textbf{0.70} & \textbf{0.50} & \textbf{0.52} &
        \textbf{0.62} & \textbf{0.83} \\
    \midrule
\end{tabular}
}
\caption{
Test-set results for the best-performing run of each model--configuration pair. In the \emph{Ord.} column, $\checkmark$ denotes instructions that explicitly require consideration of temporal event order, whereas $\times$ denotes order-neutral instructions. In the \emph{Conn.} column, $\checkmark$ indicates that query statements are joined using temporal connecting words, whereas $\times$ indicates that they are separated by spaces. ``--'' denotes that a configuration is not applicable. Standard rerankers score each query--document pair directly; OA and OP denote order-agnostic and order-preserving alignment, respectively. Bold indicates the best overall result.
}
\label{tab:main_results}
\end{table}
\begin{table}[t]
\centering
\resizebox{1\columnwidth}{!}{
\begin{tabular}{lccccccc}
    \toprule
    \textbf{Model} & \textbf{ordering} & \textbf{MRR@10} & \textbf{mAP@10} & \textbf{mAP@50} & \textbf{R@10} & \textbf{R@50} \\
    \midrule
    MiniLM-L12 & Random & 0.50 & 0.32 & 0.34 & 0.48 & 0.71 \\
    MiniLM-L12 & Original & 0.67 & 0.48 & 0.50 & 0.60 & \textbf{0.83} \\
    MiniLM-L12 & Temporal & \textbf{0.70} & \textbf{0.50} & \textbf{0.52} & \textbf{0.62} & \textbf{0.83} \\
    \midrule
    BGE-Reranker & Random & 0.44 & 0.29 & 0.32 & 0.43 & 0.76 \\
    BGE-Reranker & Original & 0.62 & 0.42 & 0.43 & 0.57 & 0.78 \\
    BGE-Reranker & Temporal & 0.65 & 0.45 & 0.47 & 0.60 & 0.80 \\
    \bottomrule
\end{tabular}}
\caption{Results by document sentences order on the test set: original, temporal (ordering the document sentences by the temporal order of the events described in them), and random (as control). Hyperparameters were selected exclusively on the development set and fixed before test-set evaluation.}
\label{tab:ordering_ablation}
\end{table}

\subsection{Results}
\label{sec:results}

Table~\ref{tab:main_results} presents the main results. We organize the analysis around our two research questions.

\paragraph{RQ1: How effectively do semantic retrievers and rerankers capture temporal relations between query events?}
Overall, the results indicate that standard retrieval and reranking models capture these constraints to a quite limited extent. In fact, among the first-stage retrievers, the purely lexical (order ignorant) BM25 achieves the strongest results, while beating all dense retrievers --- despite their semantic encoding abilities.

Making temporal progression explicit through model instructions also does not produce consistent improvements. Instructions yield no gains for the E5 and Qwen3 embedding models, nor the Qwen3-Rerank. Similarly, joining query sentences with a temporal connecting word, show no to weak improvements. Thus, neither stronger semantic modeling nor explicitly instructing compatible models to consider temporal progression reliably addresses the task.

\paragraph{RQ2: Does preserving temporal order improve reranking performance?}
The results provide clear evidence that it does. Order-preserving (OP) alignment (§\ref{sec:monotonic_alignment}) substantially outperforms both standard reranking and its order-agnostic (OA) counterpart (§\ref{sec:non_monotonic}). With MiniLM-L12, OP achieves an MRR@$10$ of 0.70 and an mAP@$10$ of 0.50, compared with 0.55 and 0.36 for OA. Similarly, with BGE-Reranker, OP significantly improves MRR@$10$ from 0.46 to 0.65 and mAP@$10$ from 0.23 to 0.45. MiniLM-L12 OP also provides the strongest results across all reported metrics, including a Recall@$50$ of 0.83. We confirm these results using paired query-level significance tests over the 50 test queries and compute 95\% confidence intervals for the corresponding pairwise performance differences. OP significantly outperforms OA and all evaluated retriever and reranker models, with $p<0.005$ in every comparison.

The OA--OP comparison more directly isolates the value of temporal structure as both approaches construct alignments from the same statement--sentence compatibility scores and only differ in the temporal constraint step. The substantial and consistent improvements obtained by OP show that query decomposition and local semantic matching alone are insufficient. In answer to RQ2, explicitly preserving the temporal progression specified by the query provides a strong additional ranking signal.

\paragraph{Effect of document sentence order.}
The sentence-ordering experiment in Table~\ref{tab:ordering_ablation} supports this interpretation. Randomizing document sentences substantially degrades performance for every sentence-pair scorer, while both the original textual order and the inferred temporal order consistently perform better. This confirms that OP is sensitive to event order and that its improvements do not arise from semantic matching alone.

Inferred temporal ordering produces small numerical improvements over the original sentence order on most metrics, but the differences are not statistically significant. This does not indicate that order is unimportant, since both meaningful orderings substantially outperform the randomized control. Rather, it suggests that ReliefWeb reports are generally written in an approximately temporal chronological manner, making their original sentence order a useful approximation of the underlying event timeline.

\section{Conclusion} 

We studied full-document retrieval for timeline sensitive queries, asking (1) whether current retrieval and reranking models capture temporal progression, and (2) whether explicitly preserving event order provides a ranking signal beyond semantic compatibility. Using a novel focused evaluation collection that separates statement-level support from temporal consistency, we find that strong semantic models provide only limited gains, while explicitly instructing compatible models to consider temporal progression yields no improvement. In answer to our first question, existing models therefore appear to poorly capture temporal constraints. For our second question, a controlled comparison between otherwise equivalent alignment models shows that order-preserving alignment significantly outperforms both its order-agnostic counterpart and all evaluated retrievers and rerankers. The degradation caused by randomizing sentence order further confirms that meaningful internal event order contributes beyond local semantic matching. Overall, our findings establish temporal progression as a valuable but yet unaddressed ranking signal, thus motivating future development of generic retrieval and reranking models that would effectively address temporal modeling as well.

\section*{Limitations}

This study is intentionally focused and uses a modest evaluation collection of 66 timeline queries, including 50 test queries. This reflects the high cost of evidence-backed temporal relevance assessment, which requires verifying multiple query statements and reconstructing their chronological relations within each document. Although paired significance tests indicate that the larger observed differences are robust, the collection provides limited statistical power for smaller effects and fine-grained subgroup analyses. We mitigate concerns about assessment reliability through repeated-assessment consistency checks, a category-pruning audit, and targeted human validation.

The benchmark is derived from ReliefWeb disaster documents, which provide public, real-world event narratives but represent only one domain.
Finally, our benchmark construction relies on LLM-assisted relevance assessment and chronological sentence ordering; although we use evidence validation and human assessment to improve reliability, errors in these steps may affect both labels and preprocessing.

\bibliography{custom}

\appendix

\section{Further Dataset Creation Details}
\label{apdx:data_create}

\subsection{Single-Event Report Filtering}
\label{apdx:multi_filter}
ReliefWeb contains many types of reports, while some describe the unfolding of a single major event and are therefore well aligned with our task, many others focus on related but distinct aspects, such as humanitarian response efforts, coordination meetings, mitigation activities, relief operations, or general situation updates covering multiple events within a region. Such reports are not necessarily irrelevant in a broad retrieval setting. However, because relevance annotation in our task is cognitively demanding --- and correspondingly expensive when performed either by human annotators or strong LLM-based annotators --- we narrow the collection construction process to reports that are more likely to fall within the scope of our formulation. In particular, we focus on reports that describe the progression of a single major event, which also constitute the most challenging and informative cases for evaluating timeline-based retrieval.

To address this, we prompt an LLM (GPT-5.2) with a structured classification prompt, instructing it to label each ReliefWeb disaster-related document as either \textit{Single}, if it describes the unfolding of a single natural disaster event, or \textit{Multiple}, if it describes multiple events or general information related to disaster response or monitoring.

\begin{table}[t]
\centering
\resizebox{0.95\columnwidth}{!}{
\begin{tabular}{l|l}
    \toprule
    \textbf{ReliefWeb Disaster Categories} & \textbf{Our Disaster Categories} \\
    \midrule
    Tropical Cyclone, Severe Local Storm, Extratropical Cyclone & Storm \\
    \midrule
    Storm Surge, Flood, Flash Flood & Flood \\
    \midrule
    Land Slide, Mud Slide & Landslide \\
    \midrule
    Fire, Wild Fire & Fire \\
    \midrule
    Drought & Drought \\
    \midrule
    Snow Avalanche & Snow Avalanche \\
    \midrule
    Cold Wave & Cold Wave \\
    \midrule
    Heat Wave & Heat Wave \\
    \midrule
    Earthquake & Earthquake \\
    \midrule
    Volcano & Volcano \\
    \midrule
    Tsunami & Tsunami \\
    \midrule
    Technological Disaster & Technological Disaster \\
    \midrule
    Insect Infestation & Insect Infestation \\
    \midrule
    Epidemic & Epidemic \\
    \midrule
    Other & Other \\
    \bottomrule
\end{tabular}}
\caption{Mapping between ReliefWeb disaster categories and our merged disaster categories.}
\label{tab:disaster-category-mapping}
\end{table}

\subsection{Disaster Categories}
\label{apdx:disaster_categ}


Exhaustively assessing all ($6,145 \times 66$) query--document pairs is impractical. 
We therefore first reduce the candidate annotation space by mapping both documents and queries to 15 predefined disaster categories derived from the ReliefWeb category set. 

\begin{figure}[!t]
\centering
\includegraphics[width=.48\textwidth]{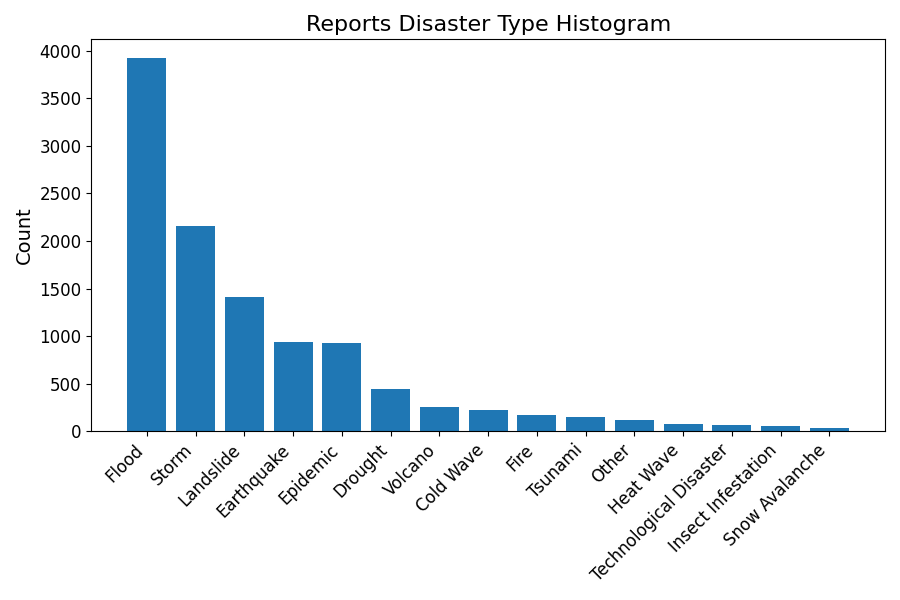}
\caption{Distribution of disaster categories over the document collection.}
\label{fig:disaster_types_report}
\end{figure}

\begin{figure}[!t]
\centering
\includegraphics[width=.48\textwidth]{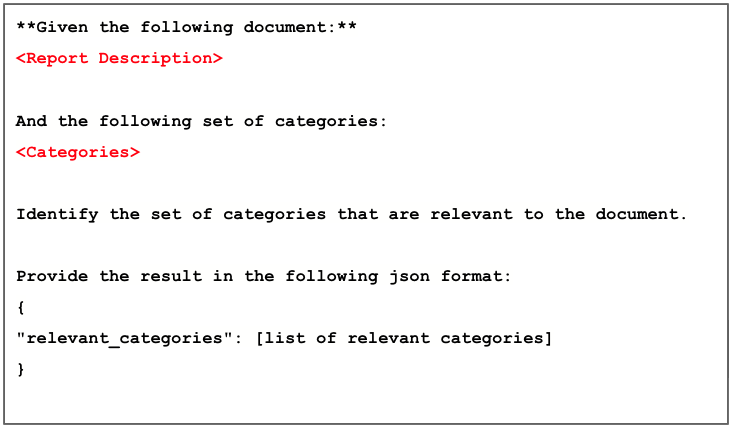}
\caption{Report categorization prompt. \textcolor{red}{<Report Description>} and \textcolor{red}{<Categories>} are replaced with the full document text and the category list, respectively.}
\label{fig:categ_prompt}
\end{figure}

ReliefWeb provides 21 disaster categories. However, we observed that these categories are too fine-grained for reliable LLM-based report categorization. To mitigate this, we created the 15 coarser-grained category set (Table~\ref{tab:disaster-category-mapping}) and prompted GPT-5.2 to identify the disaster categories described in each report from this predefined list (Figure~\ref{fig:categ_prompt}). Figure~\ref{fig:disaster_types_report} shows the distribution of reports across the final category set. A report may involve more than one disaster type; for example, a report describing an earthquake that triggered an avalanche would be assigned both \textit{Earthquake} and \textit{Avalanche}.

Using an LLM to assign categories to each document and query (see  Table~\ref{tab:disaster-category-mapping}).
We then consider only query--document pairs that share at least one disaster category, since documents describing substantially different disaster types, such as earthquakes and insect infestations, are unlikely to be relevant to the same timeline query. This category-based filtering is used only to reduce annotation effort; after annotation, all documents are treated as part of a single retrieval collection at evaluation time.
To estimate whether this category-based clustering misses relevant pairs, we manually assess 100 query--document pairs sampled from outside the shared-category candidate pool; none were found to be relevant.

\begin{figure}[!t]
\centering
\includegraphics[width=.48\textwidth]{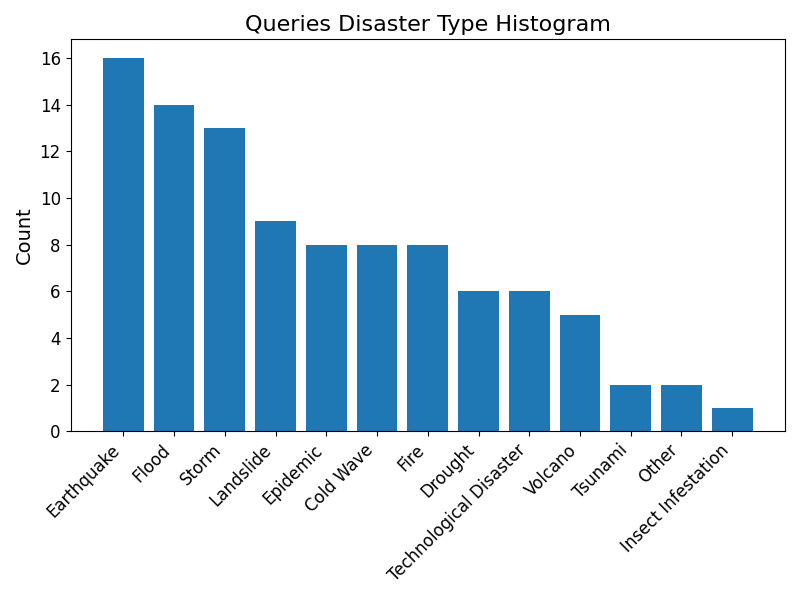}
\caption{Distribution of disaster categories over the query collection.}
\label{fig:disaster_types_query}
\end{figure}

\subsection{Temporal-order Consistency}
\label{apdx:temp_order_consist}
Further inspection of the 14\% unchanged pairs revealed that most were genuinely challenging cases in which the timeline relation involving the modified query statement could not be easily inferred, and could therefore be considered ambiguous with respect to relevance (a qualitative analysis of this experiment is presented in Appendix~\ref{apdx:additional}, Figure~\ref{fig:hard_neg1}, and \ref{fig:hard_neg2}). Under our definition of temporal consistency (§\ref{sec:task_definition}), which treats ambiguous temporal relations as consistent, the model's judgments on these cases may well be correct.

\section{Additional Experimental Details}

\subsection{Retrievals}
\label{apdx:ret_impl_details}
In our retrieval baseline experiments, we use the following implementation details. To represent each query, we concatenate all query sentences into a single text sequence, separated by spaces for control, and by ``then'' to provide temporal signals. Each document is represented by its full description text as provided in the ReliefWeb corpus. In the instruction-based model setting (E5-Large-Instr, and Qwen3-Embedding), we experiment with several instructions and reported the best one. For models whose context window cannot accommodate the full document, such as E5 and the MiniLM family, we apply a stride-based segmentation strategy, splitting each document into overlapping windows of 200 words with a stride of 100 words. We evaluate three pooling strategies over the per-segment scores: average pooling, max pooling, and weighted-position pooling, where later segments receive linearly increasing weights to reflect the chronological structure of the encoded documents.

\subsection{Rerankers}
\label{apdx:rerank_impl_details}
In our reranker baseline experiments, we use the following implementation details. In the cross-encoder settings, each query--document pair is scored jointly by taking the concatenated query text and document text as input and producing a single relevance score. For models with limited context windows, we apply the same segmentation and pooling strategies used during retrieval. In the instruction-based model setting (BGE-Reranker, and Qwen3-0.6B-Rerank), we experiment with several instructions and reported the best one.

\subsection{Instructions}
\label{apdx:instruct}
For both retriever and rerankers we investigate several time sensitive instructions: "Retrieve disaster reports that describe the same chronological sequence of events as the query.", "Find documents whose event timeline matches the order of events described in the query.", "Given a timeline query, retrieve reports where the described events occur in the same temporal order.", "Assess whether this report describes the same sequence of events as the query timeline.", "Determine if the document supports the event order described in the query, in the same chronological progression.", "Does the document contain evidence for each query event in the order they are described?",

For non-time sensitive instructions we us "Given a web search query, retrieve relevant passages that answer the query."

\subsection{Document Chronological Order Arrangement}
\label{apdx:doc_chrono_order}

As an offline preprocessing step, we split each document into sentences using NLTK \citep{bird-loper-2004-nltk}. Each sentence is assigned a unique ID, and an LLM (GPT-5.2) is prompted (see Figure~\ref{fig:order_prompt}) to return a JSON list of sentence IDs ordered according to the temporal order of the events they describe, relying on temporal expressions, such as times and dates, as well as temporal prepositions and discourse cues (before, after, later, etc.). We ask the model to output only sentence IDs, rather than regenerated sentences, to reduce hallucinations and ensure that the reordered sequence contains only sentences from the original document. If the generated output contains duplicate or missing sentence IDs, we regenerate the ordering for that document.

\subsection{Hyperparameter Tuning}
\label{apdx:hyper_tune}
In all experiments, we tune the query-gap penalty $g_q$ and the coverage weight $\alpha$ on the development set using grid search, and report final results on the test set. Specifically, we search over $g_q \in \{0, -1, \ldots, -13\}$ and $\alpha \in \{0, 0.5, \ldots, 2.0\}$. The document-skipping penalty is fixed to $g_s = 0$, reflecting the assumption that candidate documents may contain additional material that is not part of the query timeline. The best results are achieved using MiniLM-L12 (OP), with the query-gap penalty $g_q$ set to $-9$ and $\alpha$ set to $1.0$. An analysis of the effects of $g_q$ and $\alpha$ is provided in Appendix~\ref{apdx:additional}, Figures~\ref{fig:gapq_trand} and~\ref{fig:alpha_trand}.

\subsection{Order-Preserving Alignment Complexity}
\label{apdx:alignment_complexity}
For a timeline query containing $m$ statements and a document containing $n$ sentences, the order-preserving alignment runs in $O(mn)$ time after the sentence-level score matrix is computed. Thus, the alignment cost grows linearly with both the number of query statements and the number of document sentences. Since timeline queries are short, this procedure is suitable for use as a second-stage scorer over a bounded set of candidate documents. Overall, the alignment remains tractable in practice, taking 1--5 seconds per query, calculated using a single A100 GPU instance.





\subsection{Hyperparameter analysis}
\label{apdx:hyperparams}

Figures~\ref{fig:alpha_trand} and~\ref{fig:gapq_trand} analyze the sensitivity of the monotonic alignment model to the query-gap penalty $g_q$ and the coverage weight $\alpha$. 
The results show that the query-gap penalty is an important parameter: performance is strongest for moderately negative values of $g_q$, while both overly harsh penalties and overly weak penalties tend to reduce performance. 
This pattern is consistent with the role of $g_q$ in the alignment objective. 

The coverage weight $\alpha$ has a more moderate effect. 
Across the tested values, performance is relatively stable compared to the variation induced by $g_q$, suggesting that most of the ranking signal comes from the alignment path itself, while the coverage term acts as a useful regularizer. 
In particular, the best-performing configurations tend to combine a moderate query-gap penalty with a nonzero coverage weight, encouraging the model to prefer documents that both align well semantically and cover a larger portion of the query timeline.

\section{Performance by Query Length}
\label{apdx:query_length_analysis}

We further analyze retrieval performance as a function of query length; results are presented in Figure~\ref{fig:trend_analysis}. We group test queries into four bins according to the number of timeline statements: short (3--4), medium (5--6), long (7--8), and longest (9--10). Figure~\ref{fig:trend_analysis} reports MRR@10, mAP@10, mAP@50, R@10, and R@50 for BM25, the strongest standard cross-encoder reranker, and our timeline-aware reranker.

Overall, the timeline-aware reranker shows the most consistent improvement as query length increases, particularly in MRR@10, mAP@10, mAP@50, and R@10. BM25 also improves with longer queries, likely because additional timeline statements provide more lexical anchors. In contrast, the standard cross-encoder exhibits a less consistent trend and does not benefit as clearly from longer queries. These results suggest that explicit statement-level alignment can better exploit the additional constraints provided by longer timeline queries. Since the longest bin contains only a small number of queries, we treat this analysis as descriptive rather than conclusive.

\section{AI assistants}
We used AI assistants (e.g., ChatGPT) to support code formatting, phrasing suggestions, and LaTeX styling during writing. All outputs were reviewed and edited by the authors. No content was directly generated or used without human verification.

\section{Additional Illustrations}
\label{apdx:additional}

\begin{figure}[!t]
\centering
\includegraphics[width=.48\textwidth]{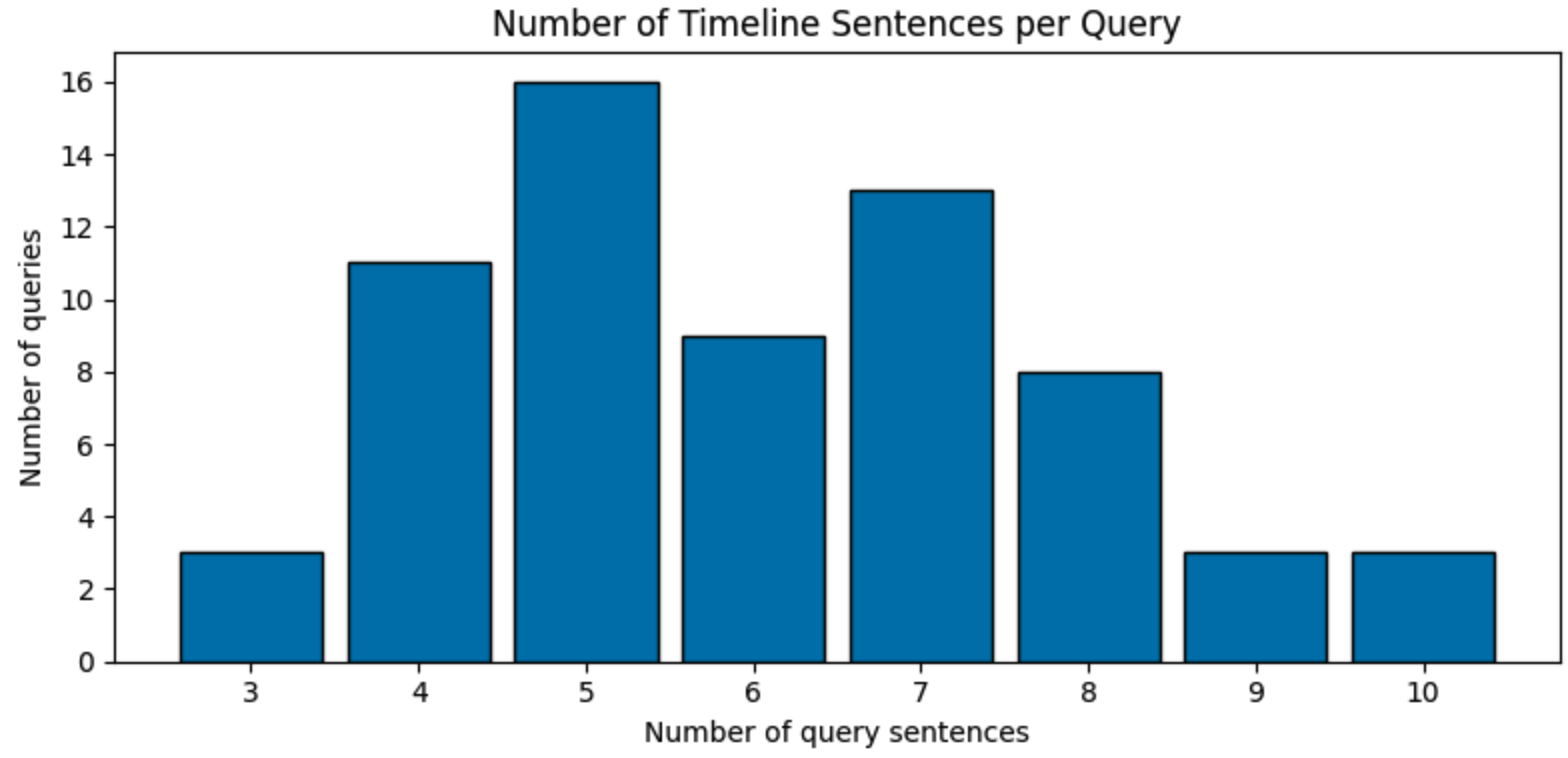}
\caption{Distribution of query lengths in the the query collection, measured by the number of event sentences per query. Most queries contain between 4 and 8 sentences, with an average of 6 event sentences.}
\label{fig:query_sents}
\end{figure}

\begin{figure}[!t]
\centering
\includegraphics[width=.48\textwidth]{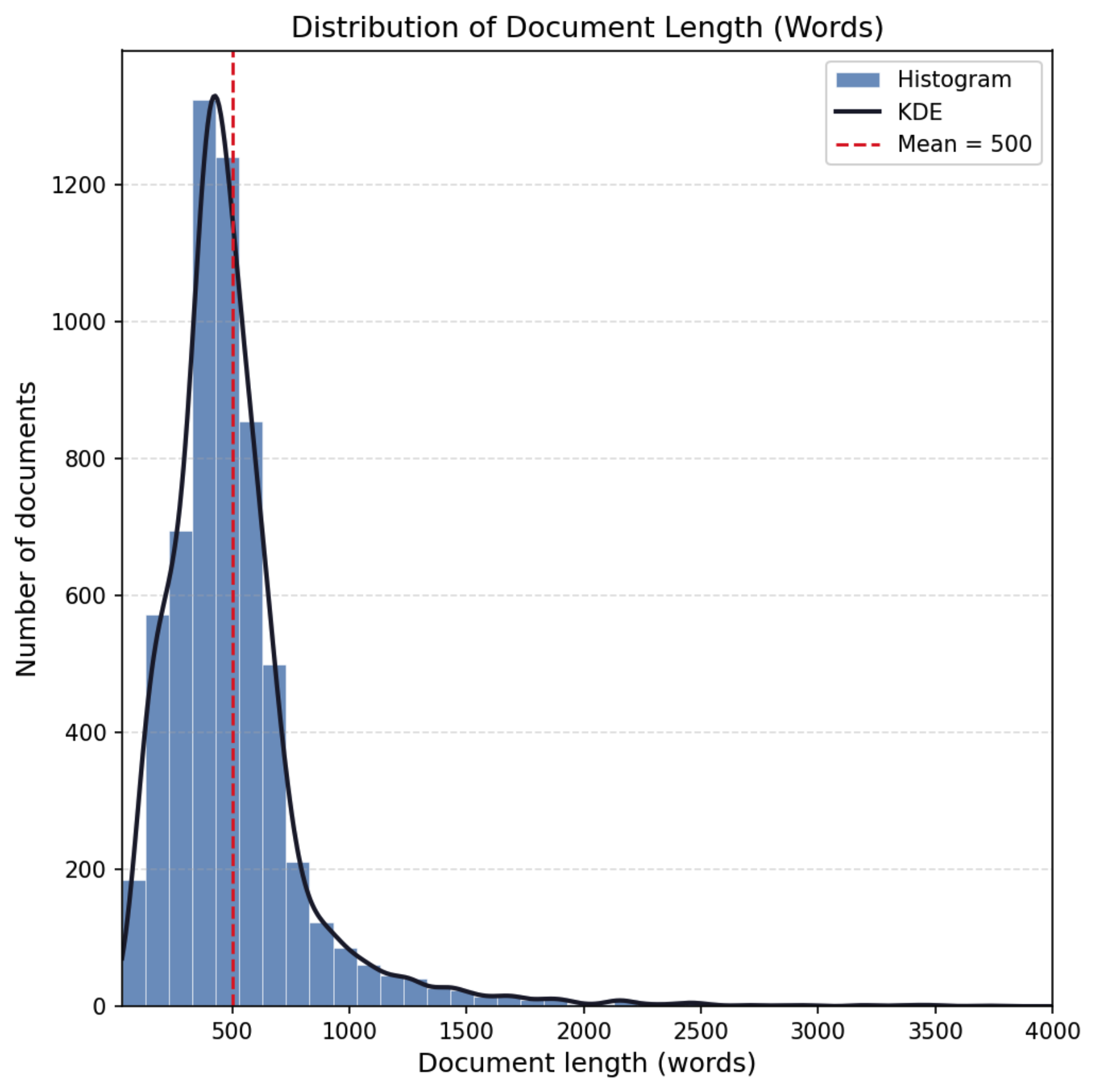}
\caption{Distribution of document lengths in words in the document collection.}
\label{fig:doc_sizes}
\end{figure}

\begin{figure*}[!t]
\centering
\includegraphics[width=\textwidth]{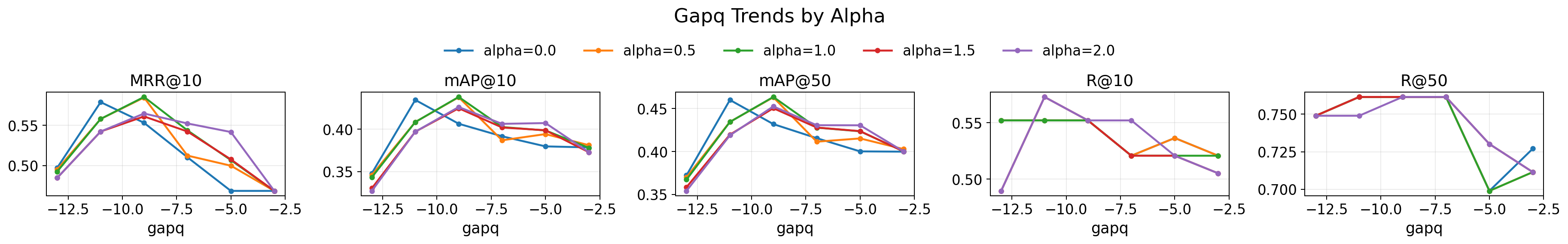}
\caption{Effect of the query-gap penalty $g_q$ across different values of the coverage weight $\alpha$.
Moderately negative gap penalties yield the strongest performance, while overly weak or overly harsh penalties reduce ranking quality.}
\label{fig:gapq_trand}
\end{figure*}

\begin{figure*}[!t]
\centering
\includegraphics[width=\textwidth]{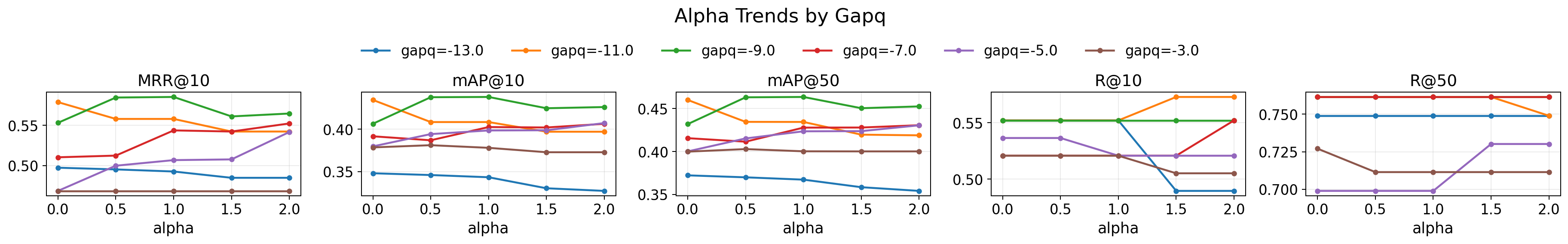}
\caption{Effect of the coverage weight $\alpha$ across different values of the query-gap penalty $g_q$.
Performance is relatively stable across $\alpha$, indicating that the alignment score provides the main ranking signal while coverage acts as a useful regularizer.}
\label{fig:alpha_trand}
\end{figure*}

\begin{figure*}[!t]
\centering
\includegraphics[width=\textwidth]{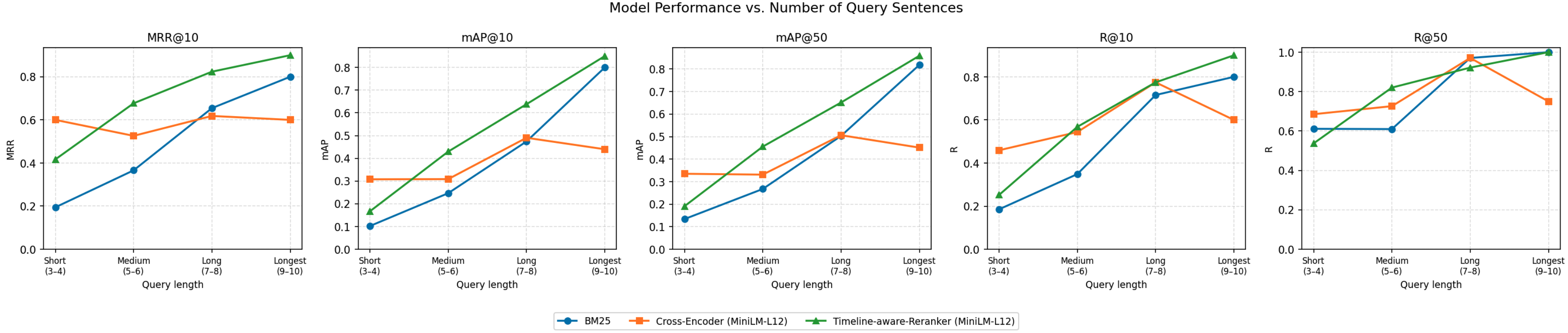}
\caption{
Model performance by query length on the test set. Queries are grouped by the number of timeline statements. The timeline-aware reranker shows the most consistent improvement as query length increases.
}
\label{fig:trend_analysis}
\end{figure*}

\begin{figure*}[!t]
\centering
\includegraphics[width=\textwidth]{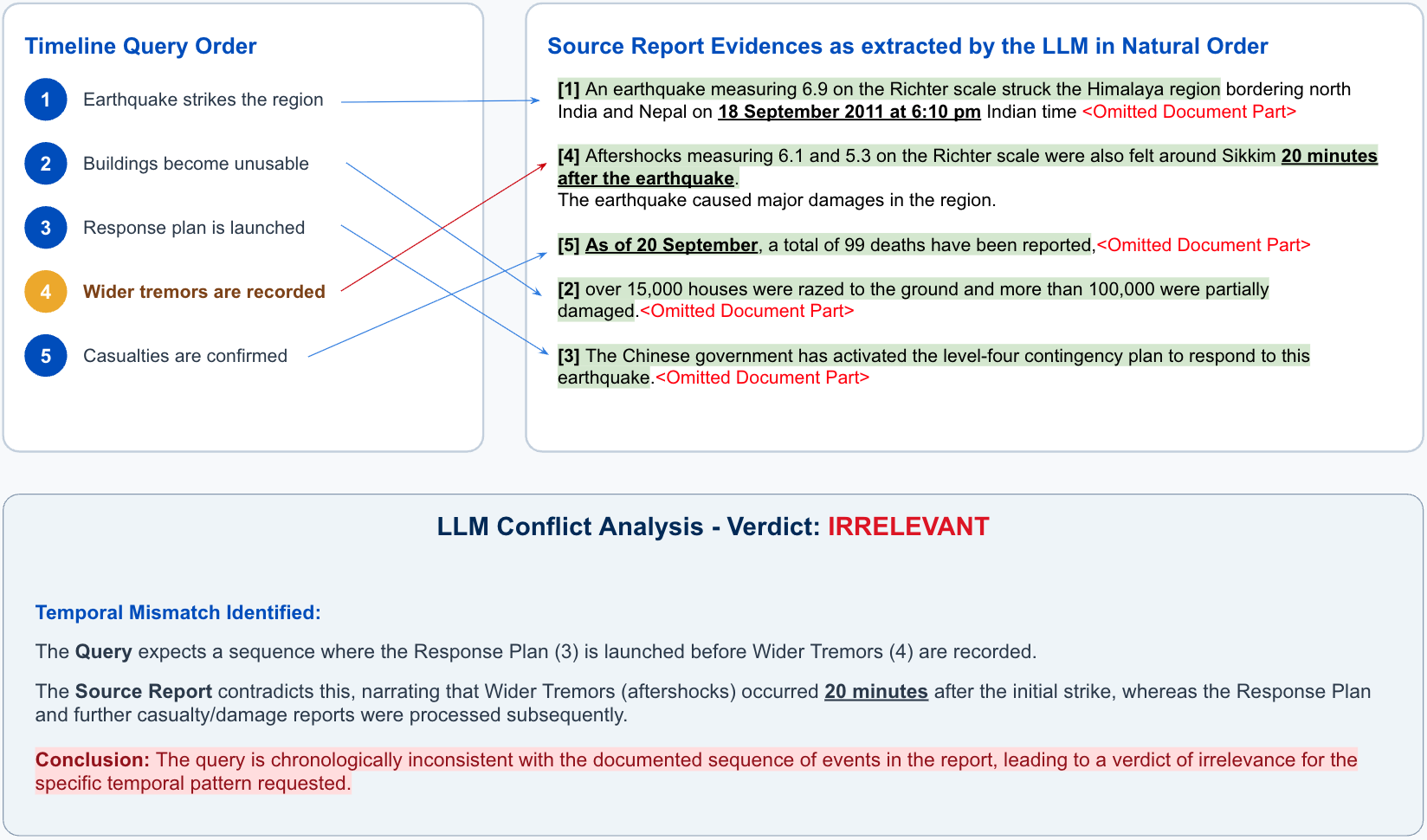}
\caption{Example of a challenging query--document pair in which GPT-5.2 identifies support for all timeline query statements, but \textbf{correctly} labels the document as \textsc{Irrelevant} because the supported events are \textbf{inconsistent} with the query timeline. The document is shown in its natural order; red \textcolor{red}{\texttt{<Omitted Document Part>}} markers indicate document sections omitted because they are irrelevant to the example and to keep the figure concise.}
\label{fig:hard_neg1}
\end{figure*}

\begin{figure*}[!t]
\centering
\includegraphics[width=\textwidth]{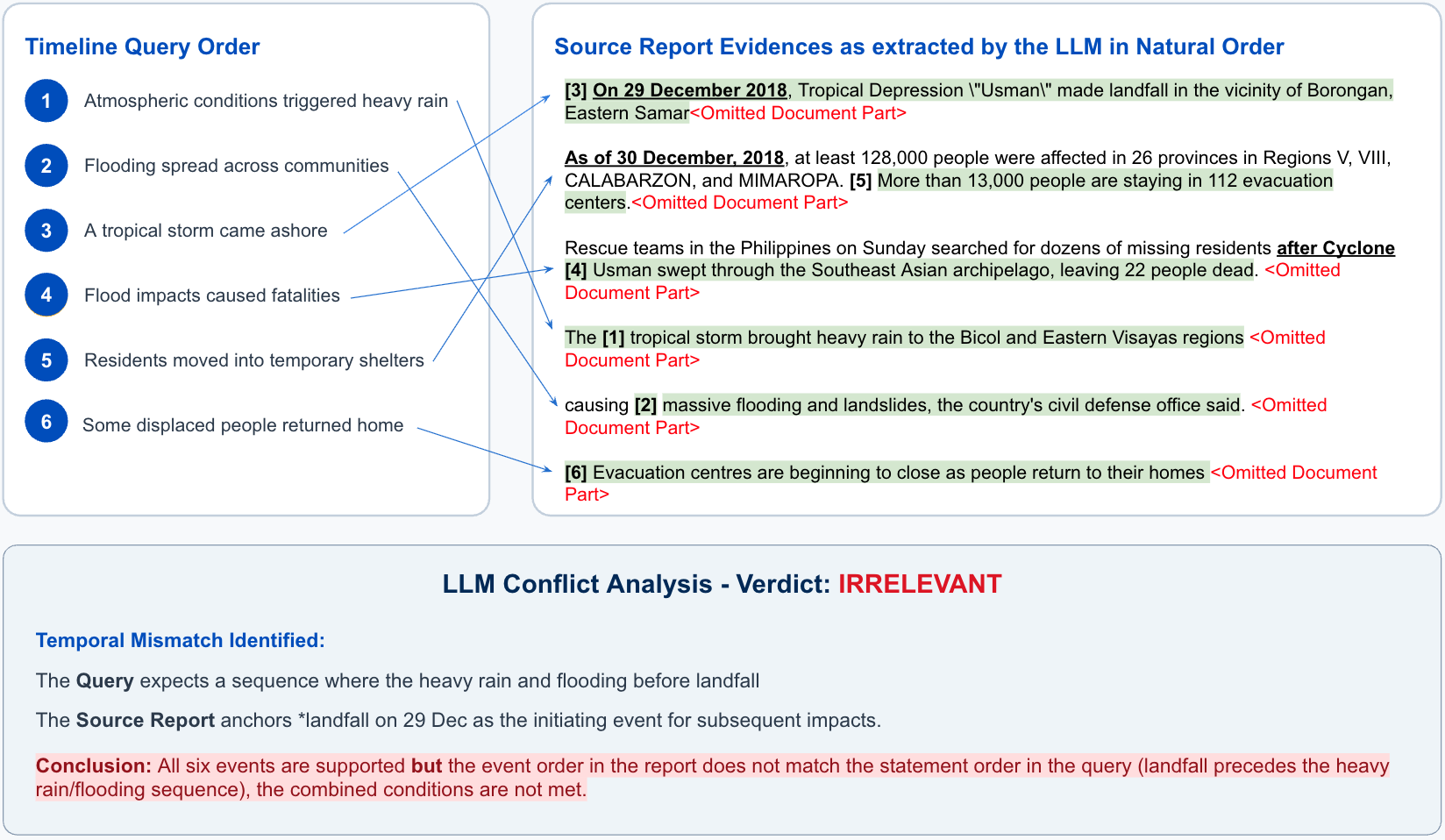}
\caption{Example of a challenging query--document pair in which GPT-5.2 identifies support for all timeline query statements, but \textbf{incorrectly} labels the document as \textsc{Irrelevant} because it hallucinates a ``landslide'' query statement (as indicated in the models ``conclusion''). In fact, the supported events are \textbf{consistent} with the query timeline. The document is shown in its natural order; red \textcolor{red}{\texttt{<Omitted Document Part>}} markers indicate document sections omitted because they are irrelevant to the example and to keep the figure concise.}
\label{fig:hard_neg2}
\end{figure*}

\begin{figure*}[!t]
\centering
\includegraphics[width=\textwidth]{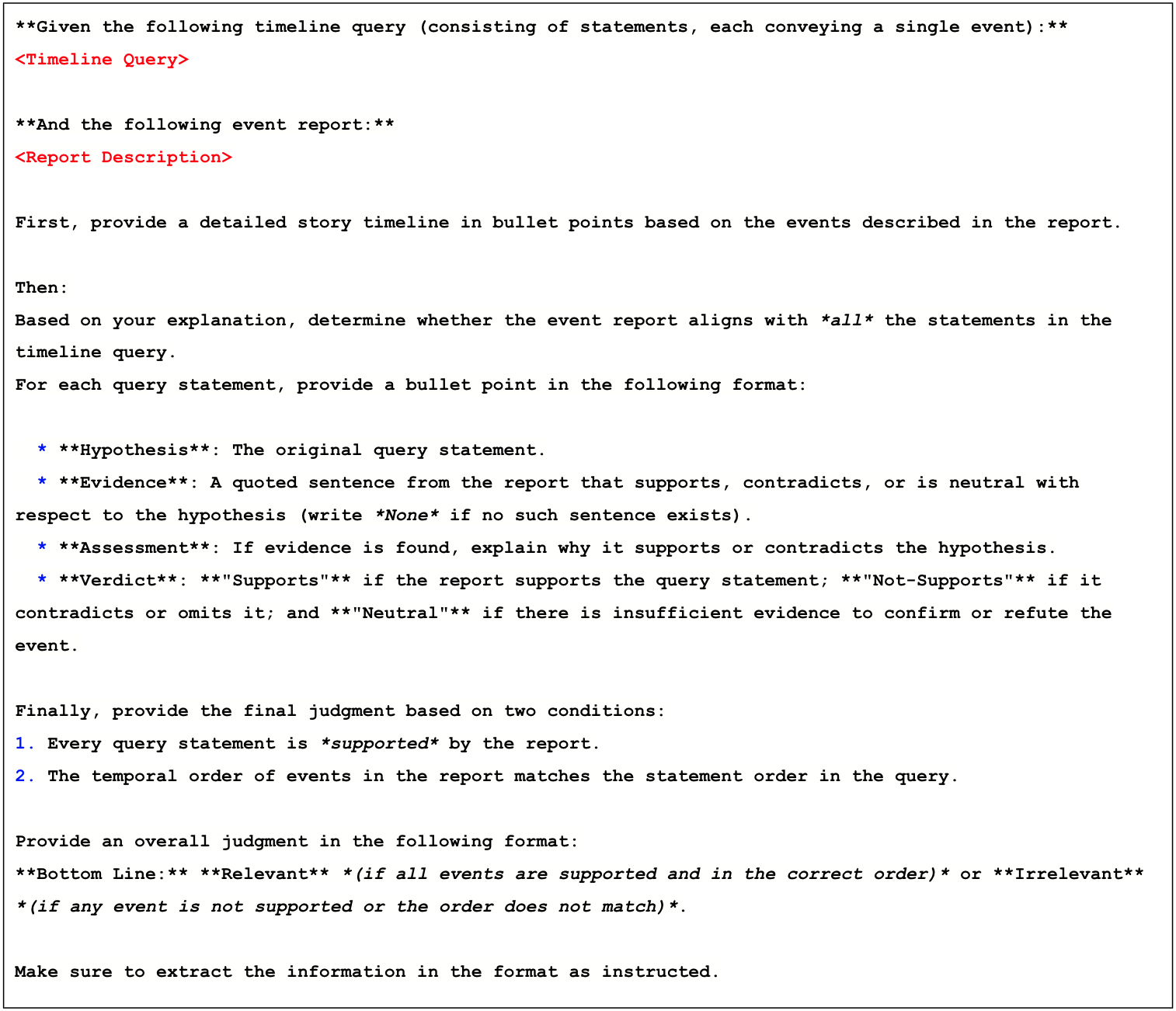}
\caption{Relevance assessment protocol prompt. \textcolor{red}{<Timeline Query>} and \textcolor{red}{<Report Description>} are replaced with the full query statements and the ReliefWeb document text, respectively.}
\label{fig:pairs_prompt}
\end{figure*}

\begin{figure*}[!t]
\centering
\includegraphics[width=\textwidth]{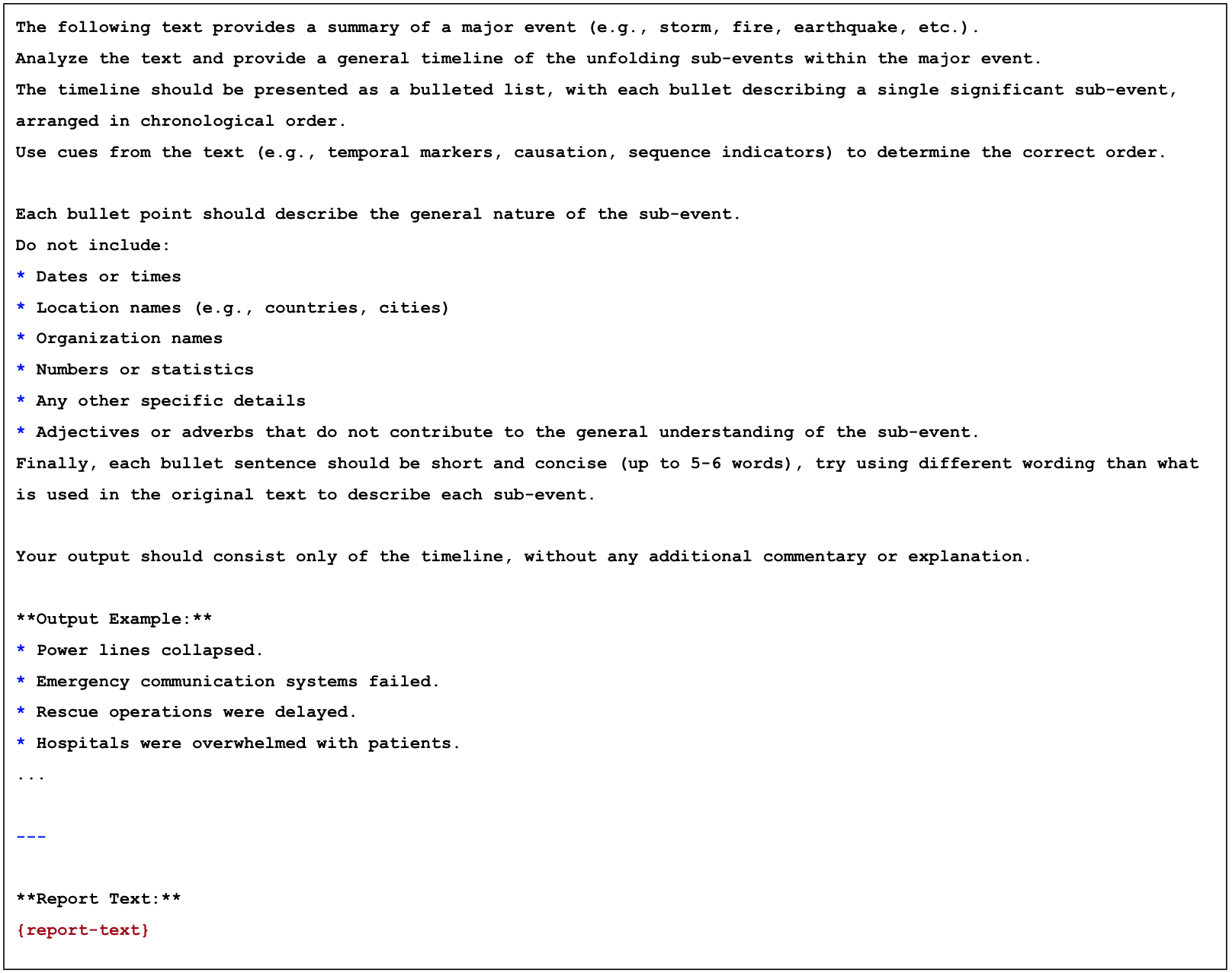}
\caption{The prompt used to generate timeline queries. \textcolor{red}{\{report-text\}} is replaced with the full ReliefWeb document text.
}
\label{fig:query_gen_prompt}
\end{figure*}

\begin{figure*}[!t]
\centering
\includegraphics[width=\textwidth]{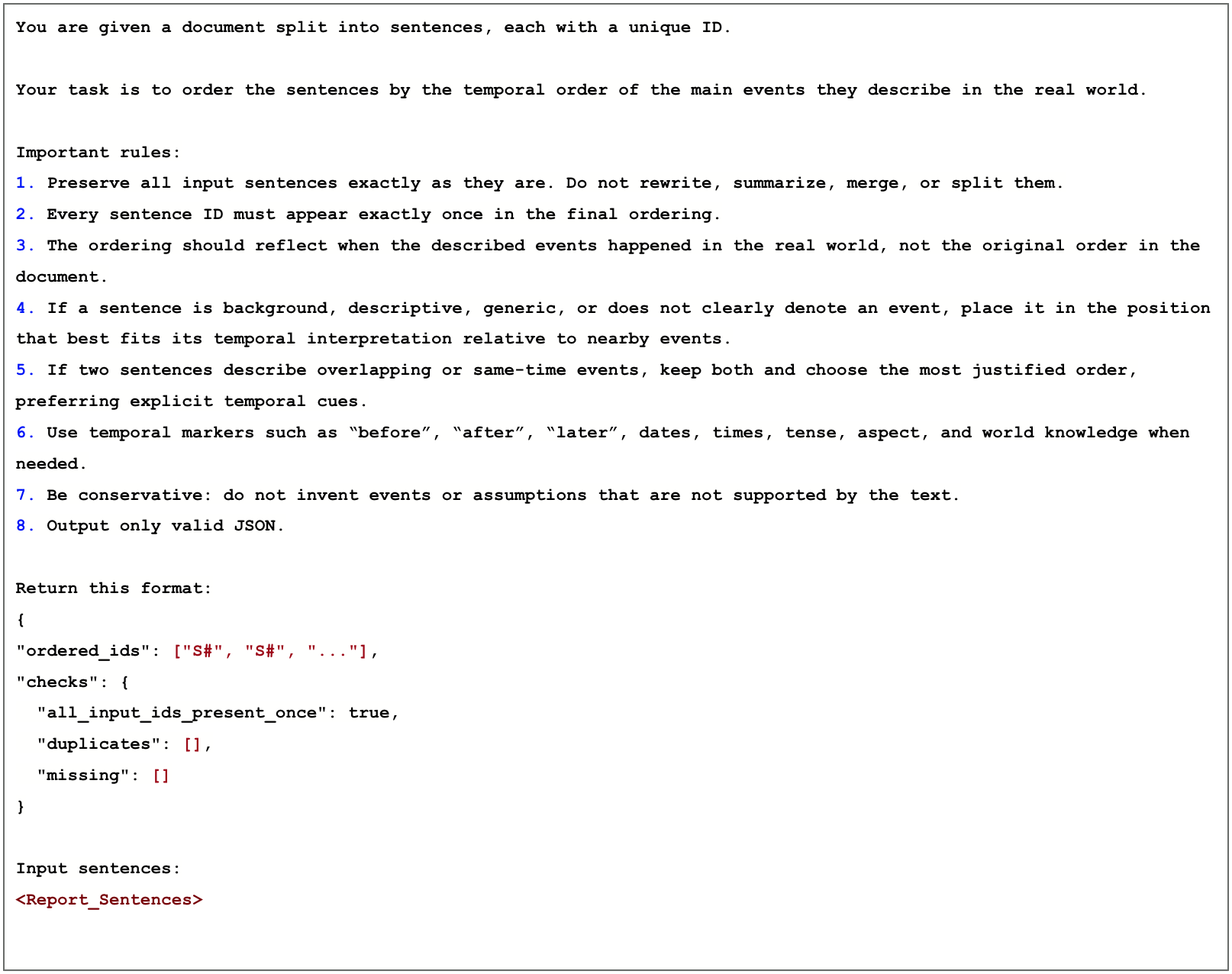}
\caption{Document chronological ordering prompt. As a preprocessing step, we split the report into sentences using NLTK and replace \textcolor{red}{<Report\_Sentences>} with the resulting sentence list.}
\label{fig:order_prompt}
\end{figure*}

This appendix provides additional qualitative examples, analysis figures, and implementation prompts that complement the main paper. Figure~\ref{fig:query_sents} shows the distribution of query lengths, Figure~\ref{fig:doc_sizes} shows the distribution of document lengths in the collection, Figures~\ref{fig:gapq_trand} and~\ref{fig:alpha_trand} analyze the sensitivity of the timeline-aware reranker to the query-gap penalty $g_q$ and the coverage weight $\alpha$. Figure~\ref{fig:trend_analysis} further reports model performance grouped by query length. We also include qualitative examples of challenging query--document pairs: Figure~\ref{fig:hard_neg1} illustrates a case where all query statements are supported but the temporal order is inconsistent, and Figure~\ref{fig:hard_neg2} shows an LLM annotation error caused by a hallucinated query statement. Table~\ref{tab:matched_query_document_examples} presents examples of the alignment paths selected by the timeline-aware reranker for top-ranked relevant documents. Finally, Figures~\ref{fig:pairs_prompt},~\ref{fig:query_gen_prompt}, and~\ref{fig:order_prompt} provide the prompts used for relevance assessment, query generation, and document chronological ordering, respectively.

\clearpage

\clearpage

\begin{table*}[t]
\centering
\small
\begin{tabularx}{\textwidth}{@{}C{0.04\textwidth}|P{0.25\textwidth}|L@{}}
\toprule
\textbf{\#} &
\multicolumn{1}{c|}{\textbf{Query Sentence}} &
\multicolumn{1}{c@{}}{\textbf{The Matched Document Sentence (ordered)}} \\
\midrule

\midrule
\multicolumn{3}{@{}c@{}}{\textbf{Query / Document Pair 1}} \\
\midrule

1 & Cold wave began impacting communities
& The cold wave struck the northern districts of the sub-Himalayan regions of Bangladesh, claiming more than 72 lives, mostly children due to hypothermia. \\
\midrule

2 & Deaths were reported in affected areas
& Hospitals in the affected districts reported an increase in the number of admissions for cold weather-related illness, particularly children, women and the elderly. \\
\midrule

3 & Dense fog disrupted daily life
& Due to this heavy fog, ferry services, port and air traffic schedules were heavily disrupted. \\

\midrule
\multicolumn{3}{@{}c@{}}{\textbf{Query / Document Pair 2}} \\
\midrule
1 & Earthquake strikes populated areas
& At 01.30 local time on 22 June, a 5.9 magnitude earthquake struck the Central Region of Afghanistan impacting Paktika and Khost provinces. \\
\midrule

2 & Deaths and injuries reported
& So far, around 920 casualties have been reported. \\
\midrule

3 & Buildings are damaged
& Further, nearly 1,500 homes have now been verified as destroyed and damaged in Giyan district, Paktika province. \\
\midrule

4 & Casualty and damage counts rise
& Additional assessments are taking place today to verify this, however, and casualties may rise further. \\
\midrule

5 & Later assessments update impacts
& Needs assessments are being completed, the final numbers of death and affected are being confirmed. \\
\midrule
\multicolumn{3}{@{}c@{}}{\textbf{Query / Document Pair 3}} \\
\midrule
1 & Initial infections were detected
& \textit{No match found}. \\
\midrule

2 & Authorities announced a new outbreak
& The Government of the Democratic Republic of the Congo (DRC) announced today that a new outbreak of Ebola virus disease is occurring in Wangata health zone, Mbandaka, in Équateur province. \\
\midrule

3 & Patient care and monitoring began
& \textit{No match found}. \\
\midrule

4 & Laboratory confirmation was conducted
& Three of these six cases have been confirmed with laboratory testing. \\
\midrule

5 & Genetic sequencing compared virus strains
& WHO, 1 Jun 2020: New genetic sequence analysis by DRC's National Institute of Biomedical Research (INRB) has found that the newly identified Ebola virus circulating in the Équateur Province in western DRC is different from the one which has infected more than 3,400 people in the eastern part of the country. \\
\midrule

6 & The outbreak was declared ended
& The outbreak in western DRC, announced on 1 June 2020, came as another Ebola outbreak in the eastern part of the country was winding down, and finally declared over on 25 June 2020. \\

\bottomrule
\end{tabularx}
\caption{Examples of timeline-aware reranker outputs for top-ranked relevant results. Each example presents a timeline query, its statements, and the document sentences matched by the alignment algorithm. Queries 1 and 2 illustrate alignments in which no query statements are skipped, while Query 3 illustrates a case where the best alignment skips two query statements.
}
\label{tab:matched_query_document_examples}
\end{table*}

\clearpage

\begin{figure*}[t]
    \centering
    \setlength{\fboxsep}{2pt}    
    \setlength{\fboxrule}{0.5pt} 
    \fbox{
        \includegraphics[width=0.80\textwidth]{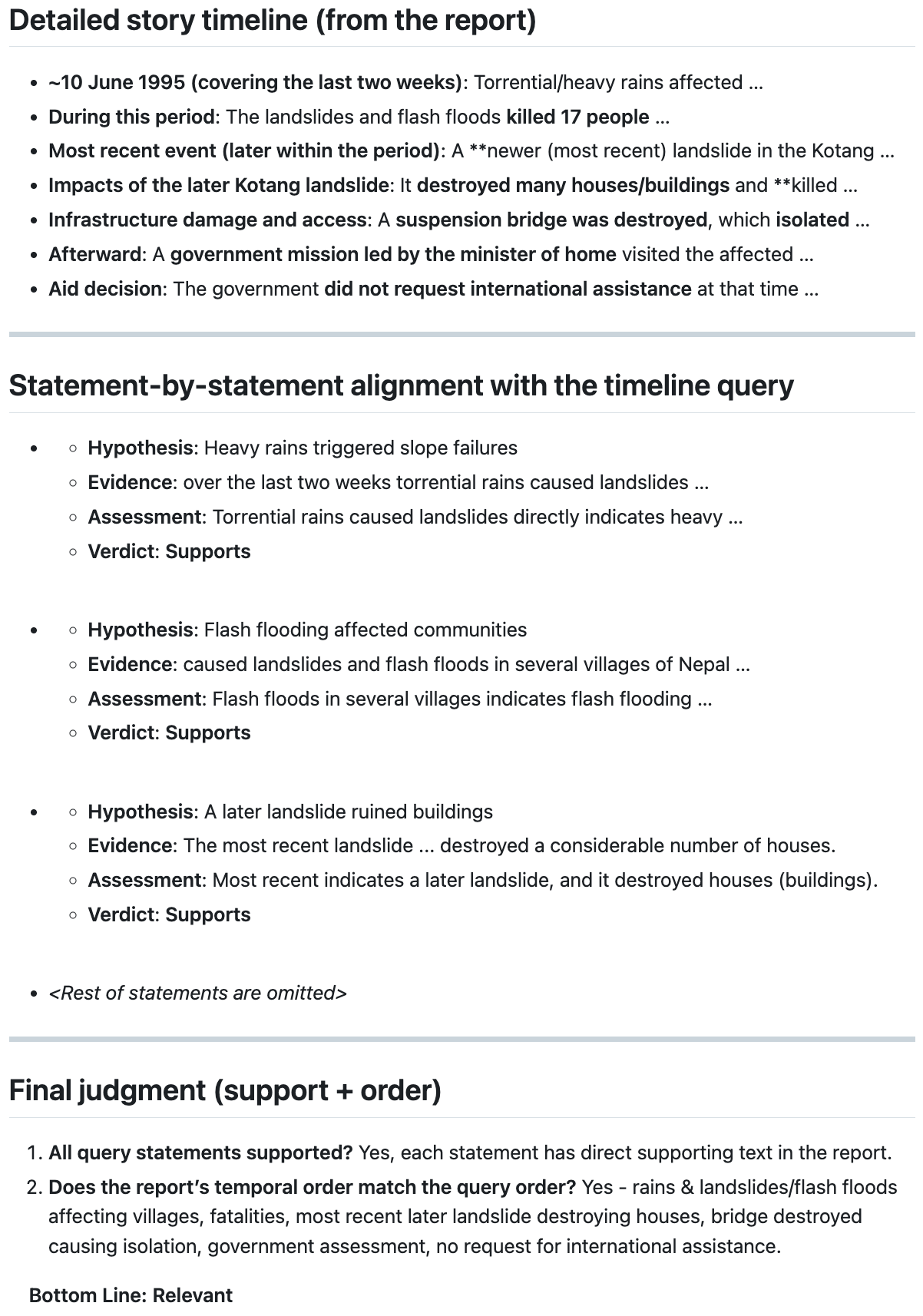}
    }
    \caption{The LLM output for the query--document relevance assessment prompt. All timeline statements and query--document alignment statements are generated, but some are trimmed in this example, as indicated by ``...'' and ``<Rest of statements are omitted>'', to ensure that the output fits within the figure. The final judgment is unchanged.}
    \label{fig:real_output}
\end{figure*}

\end{document}